\documentclass[letter,a4paper,twocolumn]{jpsj3}
\def\runauthor{
\!\!\!\!\!\!\!\!\!\!\!\!\!\!\!\!\!\!\!\!\!\!\!
Yasuhiro Yamada, Yoichi Tanaka, and Norio Kawakami}
\usepackage{txfonts}
\usepackage{bm}
\usepackage[dvips]{color}
\usepackage{ulem}

\title{Enhanced Andreev Tunneling via the Kondo Resonance in a Quantum Dot at Finite Bias}

\author{Yasuhiro Yamada\thanks{E-mail address: y.yamada@scphys.kyoto-u.ac.jp}, Yoichi Tanaka$^{1}$, and Norio Kawakami}
\inst{Department of Physics, Kyoto University, Kyoto 606-8502, Japan\\
$^{1}$Condensed Matter Theory Laboratory, RIKEN, Wako 351-0198, Japan}

\abst{We study the nonequilibrium transport through a quantum dot coupled to normal and superconducting leads. We use the modified second-order perturbation theory to calculate the differential conductance and the local density of states at the quantum dot. In the strong but finite Coulomb interaction regime, the differential conductance shows an anomalous peak not at a zero bias voltage but at a finite bias voltage. We also observe an additional Kondo resonance besides the normal one in the local density of states, where the former is caused by nonequilibrium Andreev tunneling via the normal Kondo resonance. We explain that this specific Andreev tunneling gives rise to the anomalous peak in the differential conductance. Since the Andreev tunneling via the Kondo resonance is suppressed with increasing temperature, the anomalous peak in the differential conductance disappears at high temperatures.

}

\begin{document}
\maketitle

In recent decades, the Kondo effect has received renewed interest in the context of mesoscopic systems fabricated with advanced nanotechnology. Particularly, the nonequilibrium transport through a quantum dot (QD)~\cite{reviews} coupled to two normal leads has been the subject of intense study, because interesting many-body effects are presumed to exist under nonequilibrium steady-state conditions. Furthermore, if such a QD is coupled to superconducting and normal leads (the N-QD-S system), more fascinating phenomena, such as the Andreev reflection, can be observed. 

In order to elucidate the interplay between the Andreev reflection and the Kondo effect, extensive experimental~\cite{Graber, Deacon} and theoretical~\cite{Fazio, Schwab, Cho, Clerk, Cuevas, Sun01, Krawiec, Tanaka, Domanski} studies of N-QD-S systems have been conducted. Although some theoretical studies have already addressed the nonequilibrium transport in N-QD-S systems focusing on the influence of the Kondo effect~\cite{Fazio, Schwab, Cho, Sun01, Krawiec, Domanski}, there remain several important issues to be resolved. For example, there is still controversy about the zero-bias anomaly in the differential conductance: some studies have shown the absence~\cite{Fazio, Schwab, Krawiec} (existence~\cite{Cho,Domanski}) of a zero-bias anomaly in the Kondo limit. Moreover, most theoretical studies of the nonequilibrium transport have been concerned with the limit of infinitely strong Coulomb interaction at the QD.~\cite{Fazio, Schwab, Cho, Krawiec} However, since sufficiently large Coulomb interactions exclude the superconducting correlations at the QD, no interplay between the Andreev reflection and the Kondo effect may be observed under such an assumption. Thus, it is necessary to consider the intermediate interaction regime.

In order to clarify the interplay between the Andreev reflection and the Kondo effect, we investigate the nonequilibrium transport in the N-QD-S system over a broad range of Coulomb interactions. For this purpose, we extend the modified second-order perturbation theory~\cite{Cuevas, Kajueter}, which gives the self-energy interpolating between the weak and strong coupling limits, to a nonequilibrium case by employing the Keldysh technique. This method enables us to study the nonequilibrium system in the intermediate interaction regime. From the results of the differential conductance and the local density of states (LDOS) at the QD, we elucidate how the Kondo effect affects the Andreev transport at a finite bias. In particular, a unique phenomenon in transport is observed in the intermediate regime: the nonequilibrium Andreev tunneling via the normal Kondo resonance causes an anomalous peak in the differential conductance not at a zero bias voltage but at a finite bias voltage. Its origin is discussed in detail from the viewpoint of the competition/cooperation between the Kondo and superconducting correlations. Note that since the anomalous peak appears only in the intermediate regime and seems to disappear in the limit of the infinitely strong Coulomb interaction, the results obtained here are consistent with most of the previous ones reported in this extreme limit.~\cite{Fazio, Schwab, Krawiec}

We describe the N-QD-S system in terms of an Anderson impurity coupled to both the normal metal and the BCS superconductor, $H=H_\mathrm{QD}+H_\mathrm{N}+H_\mathrm{S}+H_\mathrm{TN}+H_\mathrm{TS}$, where $H_\mathrm{QD}=\epsilon_\mathrm{d}\sum_{\sigma} d_{\sigma}^{\dagger} d_{\sigma}^{} + Ud_{\uparrow}^{\dagger} d_{\uparrow}^{} d_{\downarrow}^{\dagger} d_{\downarrow}^{}$ represents the Hamiltonian for the QD. $H_\mathrm{N} = \sum_{\sigma}(\epsilon_{k}-\mu_N)c_{N k \sigma}^{\dagger}c_{N k \sigma}^{}$ and $H_\mathrm{S} = \sum_{k\sigma}(\epsilon_{k}-\mu_S)c_{S k \sigma}^{\dagger}c_{S k \sigma}^{} + \sum_{k}(\Delta{}c_{S k \downarrow}^{\dagger}c_{S -k \uparrow}^{\dagger} + \mathrm{H.c.})$ denote the normal lead and superconducting lead, respectively ($\Delta$ is the superconducting gap). We assume that the bias voltage $V$ is applied only to the normal lead: $\mu_{N}=eV$ and $\mu_{S}=0$. The tunneling Hamiltonians between the QD and the two leads are $H_\mathrm{TN} = \sum_{k \sigma}(t_\mathrm{N}c_{N k \sigma}^{\dagger}d_{\sigma}^{} + \mathrm{H.c.})$ and $H_\mathrm{TS} = \sum_{k \sigma}(t_\mathrm{S}c_{S k \sigma}^{\dagger}d_{\sigma}^{} + \mathrm{H.c.})$, where $t_{N (S)}$ is the tunneling amplitude between the QD and the normal (superconducting) lead. We consider the wide-band limit of electrons in the leads, in which $\Gamma_\mathrm{N (S)}=\pi t_\mathrm{N (S)}^{2} \rho_\mathrm{N (S)}$ defined at $\Delta=0$ represents the resonance strength between the QD and the normal (superconducting) lead.

The time-dependent current flowing from the normal (superconducting) lead into the QD, $\hat{I}_{N(S)}$, is described by $\hat{I}_{N(S)}(t) = (ie/\hbar)\sum_{k\sigma}t_{N(S)} c^{\dagger}_{N(S)k\sigma}(t)d_{\sigma}(t) + \mathrm{H.c.}$ Since we consider the nonequilibrium steady-state transport, the expectation values of the current $\langle\hat{I}_{N(S)}\rangle$ is time-independent. Using the standard Keldysh Green's function formalism, $\langle\hat{I}_{N(S)}\rangle$ can be expressed in terms of the local retarded and lesser Green's functions at the QD, $\bm{G}^{<}(\omega)$ and $\bm{G}^{r}(\omega)$.~\cite{Fazio, Sun00} The bold font represents a $2\times 2$ Nambu matrix. These Green's functions include the self-energies due to the Coulomb interaction $U$ in addition to those due to the coupling between the QD and the leads. In order to obtain the self-energies for $U$, we use the modified second-order perturbation theory~\cite{Cuevas, Kajueter} in the Keldysh Green's function formalism which is a generalization of Cuevas's approach~\cite{Cuevas} in the equilibrium N-QD-S system to the system under nonequilibrium conditions. Following ref. \citen{Cuevas}, the modified retarded self-energy $\widetilde{\bm{\Sigma}}_{U,2nd}^{r}$ is given by a functional of the bare second-order self-energy $\bm{\Sigma}_{U,2nd}^{r}(\omega)$ such that $\widetilde{\bm{\Sigma}}_{U,2nd}^{r}$ has an appropriate form in both the atomic limit, $\Gamma_{N,S}/U \to 0$, and weak-interaction limit, $U/\Gamma_{N,S} \to 0$. In a similar manner, we obtain the modified lesser self-energy, $\widetilde{\bm{\Sigma}}_{U,2nd}^{<}$.~\cite{Cuevas}

We can calculate the bare second-order self-energies $\bm{\Sigma}_{U,2nd}^{r}$ and $\bm{\Sigma}_{U,2nd}^{<}$ using the perturbation theory with respect to $U$ in the standard  Keldysh Green's function formalism. The only point to note in this approach is that the Hamiltonian for the QD is replaced by the following modified one when we calculate the self-energy: $H_\text{QD}^{\textrm{eff}} = \bar{\epsilon}_{d}\sum_{\sigma} d_{\sigma}^{\dagger} d_{\sigma}+(\bar{\Delta}_{d} d_{\uparrow}^{\dagger}d_{\downarrow}^{\dagger} + {\rm H.c.}) + \bar{\lambda}(\hat{I}_{N}+\hat{I}_{S})$,
where the effective parameters $\bar{\epsilon}_{d}$ and $\bar{\Delta}_{d}$ are introduced to ensure that the self-energies have the correct forms in the atomic limit~\cite{Cuevas, Kajueter}. In addition, we introduce the source term $\bar{\lambda}$ coupled to the current operator in order to conserve the current, because the simple application of the second-order self-energy in the N-QD-S system without the source term may break the current conservation law: $\langle \hat{I}_{N} \rangle + \langle \hat{I}_{S} \rangle \neq 0$.~\cite{Hershfield}
The effective parameters $\bar{\epsilon}_{d}$, $\bar{\Delta}_{d}$, and $\bar{\lambda}$ are then fixed under the following conditions, which respectively indicate the consistency of the electron number, the effective counter gap to the proximity effect at the QD, and the current conservation: $\langle \hat{n} \rangle=\langle \hat{n} \rangle_{0}$, $\bar{\Delta}_{d}=[\bm{\Sigma}_{U}^{r}(0)]_{12}$ and $\langle \hat{I}_{N} \rangle + \langle \hat{I}_{S} \rangle=0$. In the first equality, $\langle \hat{n} \rangle$ and $\langle \hat{n} \rangle_{0}$ are the expectation values of the electron number at the QD with the full Hamiltonian and with the modified one.

\begin{figure}[t]
\centering
\includegraphics[width=7cm,clip]{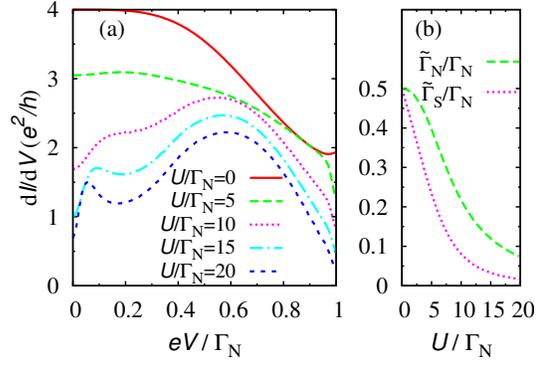}% Here is how to import EPS art
\caption{\label{fig:Udep} (Color online) (a) Differential conductance as a function of bias voltage $V$ for several values of the Coulomb interaction $U$: $\epsilon_{d}/U=-0.5$, $\Delta/\Gamma_{N}=1$, $\Gamma_{S}/\Gamma_{N}=1$, and $k_{B}T/\Gamma_{N}=0.01$. (b) Renormalized parameters $\tilde{\Gamma}_{N}$ and $\tilde{\Gamma}_{S}$ as a function of the Coulomb interaction $U$: $\epsilon_{d}/U=-0.5$, $\Delta/\Gamma_{N}=1$, $\Gamma_{S}/\Gamma_{N}=1$, $eV/\Gamma_{N}=0$, and $k_{B}T/\Gamma_{N}=0$.}
\end{figure}

First, we discuss how the differential conductance $\mathrm{d}I/\mathrm{d}V$ is affected by the Coulomb interaction. Figure \ref{fig:Udep}(a) shows the differential conductance as a function of the bias voltage $V$ for several values of $U$ with the symmetric coupling $\Gamma_{S}/\Gamma_{N}=1$. We only show $\mathrm{d}I/\mathrm{d}V$ for a positive bias voltage since it is an even function of $V$ in this case ($\epsilon_{d}+U/2=\mu_{S}=0$). For $U=0$, the zero-bias conductance takes its maximum value $4e^2/h$, where the Andreev tunneling between the QD and the superconducting lead is balanced by the normal electron tunneling between the QD and the normal lead. This balance is ensured by the condition of $\Gamma_{S}/\Gamma_{N}=1$ with $U=0$. With the enhancement of the Coulomb interaction, the differential conductance decreases gradually in the entire voltage range, because the Coulomb interaction suppresses both the Andreev tunneling and the normal electron tunneling, but in different ways: the interaction makes the former much weaker than the latter. This is evident in renormalized parameters, $\tilde{\Gamma}_{N}$ and $\tilde{\Gamma}_{S}$, which are defined for $V=0$ and $T=0$ as
\begin{eqnarray}
\tilde{\Gamma}_{N}=z\Gamma_{N},&& \tilde{\Gamma}_{S}=z\left(\Gamma_{S}+\left[\mathrm{Re}\bm{\Sigma}^{r}_{U}(\omega)\right]_{12}\right),
\end{eqnarray}
where $z$ is the renormalization factor $z=(1+\Gamma_{S}/\Delta- \mathrm{d} \left[\mathrm{Re}\bm{\Sigma}^{r}_{U}(\omega)\right]_{11}/\mathrm{d} \omega \big|_{\omega=0})^{-1}$. Figure~\ref{fig:Udep}(b) shows the $U$ dependence of these renormalized parameters. It is seen that $\tilde{\Gamma}_{S}$  decreases more rapidly than $\tilde{\Gamma}_{N}$ with an increase in $U$. For $\tilde{\Gamma}_{S} < \tilde{\Gamma}_{N}$, the coupling between the QD and the superconducting lead effectively becomes weaker than the coupling between the QD and the normal lead, making the Kondo singlet dominant at the QD. As a result, the system loses the above subtle balance in tunneling processes, leading to a global reduction in conductance. Note here that the differential conductance markedly changes its voltage dependence with an increase in $U$ (Fig. \ref{fig:Udep}(a)). In particular, the monotonic curve of the differential conductance for $U/\Gamma_{N}=0$ gradually develops a double-peak structure, as seen for $U/\Gamma_{N}=10$. As $U$ further increases, the two peaks become more prominent. The peak in the vicinity of $V=0$ becomes sharper and gradually approaches a zero-bias voltage: e.g., it is located at $eV/\Gamma_{N} \simeq 0.05$ for $U/\Gamma_{N}=20$. The anomalous sharp peak is commonly observed when the Kondo effect is dominant at the QD ($\tilde{\Gamma}_{S}/\tilde{\Gamma}_{N} < 1$ and $U/\Gamma_{N} \gg 1$) though its height decreases with increasing $U$. On the other hand, the broad peak at $eV/\Gamma_{N} \simeq 0.6$ is insensitive to the increase in $U$. The above novel characteristic feature is one of the main findings of this study, which is quite different from an ordinary zero-bias anomaly in a QD coupled to two normal leads.~\cite{reviews} Note that with increasing $U$, the anomalous peak gets close to the zero-bias voltage but decreases in height. Eventually, the anomalous peak seems to disappear in the limit of $U\to\infty$, which is consistent with the absence of the zero-bias anomaly in refs. \citen{Fazio}, \citen{Schwab}, \citen{Krawiec}, and \citen{Clerk} where an infinitely strong Coulomb interaction is assumed. The monotonic decrease in the zero-bias conductance with increasing $U$ is also consistent with the results of equilibrium studies.~\cite{Cuevas, Tanaka}
%%%%%%%%%%%%%%%%%%%%%%
%%%%%%%%%%%%%%%%%%%%%%

\begin{figure}[t]
\centering
\includegraphics[width=7cm,clip]{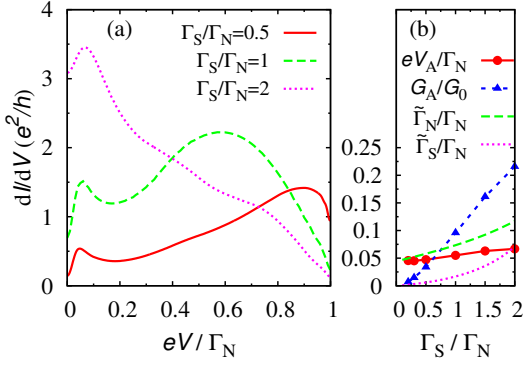}% Here is how to import EPS art
\caption{\label{fig:Gsdep} (Color online) (a) Differential conductance as a function of bias voltage $V$ for several values of $\Gamma_{S}$ and $U/\Gamma_{N}=20$. The other parameters are the same as those in Fig.~\ref{fig:Udep}(a). (b) Anomalous sharp peak position $V_{A}$ and its height $G_{A}$ as a function of $\Gamma_{S}$. The parameters are the same as those in (a). Here, $G_{0}=16e^2/h$. The renormalized parameters defined for $V=0$ and $T=0$, $\tilde{\Gamma}_{N}$ and $\tilde{\Gamma}_{S}$, are also shown.}
\end{figure}

We also show the differential conductance in the cases with asymmetric couplings ($\Gamma_{S}/\Gamma_{N} \neq 1$) and a strong Coulomb interaction $U/\Gamma_{N}=20$ in Fig.~\ref{fig:Gsdep}(a). For  $\Gamma_{S}/\Gamma_{N}>1$ ($<1$), the sharp peak near the zero-bias voltage increases (decreases) in height and changes its position toward a slightly higher (lower) bias voltage. To be more specific, let us denote the position and height of the sharp peak as $V_{A}$ and $G_{A}$, respectively, which are plotted in Fig.~\ref{fig:Gsdep}(b) together with the renormalized parameters. A significant feature is that the peak position ($eV_{A}$) is approximately given by $\tilde{\Gamma}_{N}$ for small $\Gamma_{S}$'s, where  $\tilde{\Gamma}_{N}$ corresponds roughly to the Kondo temperature. Also note that the peak height $G_{A}$ approaches zero for $\Gamma_{S} \to 0$ because the Andreev tunneling completely disappears in that limit. These facts indicate that the origin of this characteristic peak is related to both the Kondo effect and the proximity effect. Note that the other broad peak is also affected by the change in $\Gamma_{S}$. For smaller $\Gamma_{S}/\Gamma_{N}$ values, the broad peak at $eV/\Gamma_{N} \simeq 0.6$ in the symmetric case ($\Gamma_{S}/\Gamma_{N} = 1$) shifts toward the gap edge, $eV/\Gamma_{N} = 1$. In contrast, for larger $\Gamma_{S}/\Gamma_{N}$ values, the broad peak shifts in the opposite direction and merges with the sharp peak near the zero-bias voltage. All the above characteristic features of the conductance will be clearly explained below in terms of the interplay between the Kondo/proximity effects under nonequilibrium conditions.

%%%%%%%%%%%%%%%%%%%%%%
%%%%%%%%%%%%%%%%%%%%%%

\begin{figure}[t]
\centering
\includegraphics[width=7cm,clip]{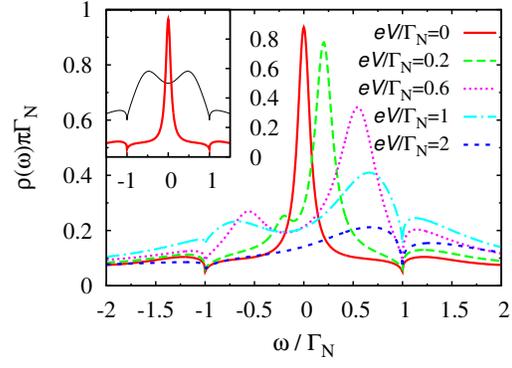}% Here is how to import EPS art
\caption{\label{fig:dos} (Color online) LDOS at the QD for several values of bias voltage $V$. We set $\epsilon_{d}/U=-0.5$, $U/\Gamma_{N}=20$, $\Gamma_{S}/\Gamma_{N}=1$, $\Delta/\Gamma_{N}=1$, and $k_{B}T/\Gamma_{N}=0.01$. The LDOS for $eV/\Gamma_{N}=0$ is shown in the inset in comparison with the one for the noninteracting case (thin solid line).}
\end{figure}

Let us now discuss how the LDOS at the QD, $\rho(\omega)$, in the strong Coulomb interaction regime changes its profile as bias voltage increases. The computed results are shown in Fig. \ref{fig:dos}. Let us first look at the LDOS in the equilibrium state ($V=0$) for $U/\Gamma_{N}=20$, which is compared with that in the case of $U/\Gamma_{N}=0 $ shown in the inset. The two resonances emerging inside the superconducting gap for $U/\Gamma_{N}=0$ represent the Andreev resonant states, which are approximately described by the superposition of two Lorentzians with renormalized parameters: they are located at $\omega = \pm \tilde{\Gamma}_{S}$ with a width $\tilde{\Gamma}_{N}$. It is seen that the Andreev resonance for $U/\Gamma_{N}=0$ is strongly renormalized and merged into a single sharp Kondo resonance at the Fermi level for $U/\Gamma_{N}=20$ because $\tilde{\Gamma}_{S}$ decreases more rapidly than $\tilde{\Gamma}_{N}$ (see Fig.~\ref{fig:Udep}(b)). The resulting Kondo resonance is sensitive to the change in bias voltage, so that the LDOS for $U/\Gamma_{N}=20$ significantly alters its shape with an increase in $V$. Indeed, as bias voltage increases, the position of the Kondo resonance shifts following the Fermi level of the normal lead $\mu_{N}$: e.g.,  for $\mu_{N}/\Gamma_{N}=eV/\Gamma_{N}=0.2$, the peak position is located at $\omega/\Gamma_{N} \simeq 0.2$. This implies that the ordinary Kondo resonance is formed by conduction electrons in the normal lead. A noteworthy feature for $eV/\Gamma_{N}=0.2$ is that another resonance develops, though small, near the counterposition of the Kondo resonance $\omega \simeq -\mu_{N}$. This additional Kondo resonance was previously observed by Sun {\it et al.}~\cite{Sun01}, but has not been discussed in detail, particularly as regards its physical relevance to the transport properties. We will address this issue below and demonstrate that it indeed provides a source of the marked change in nonequilibrium transport properties. For $eV/\Gamma_{N}=0.6$, the peak value of the additional Kondo resonance is slightly increased while the normal Kondo resonance is suppressed, leading to a broad two-peak structure, which is analogous to that in the $U/\Gamma_{N}=0$ case. The two-peak structure therefore indicates that the superconducting pairing state, which is strongly suppressed by the large Coulomb interaction at $V=0$, is revived and becomes dominant over the Kondo singlet state. With further increase in bias voltage ($eV/\Gamma_{N}=1,2$), however, the Andreev resonance is suppressed again since the applied bias voltage also makes the superconducting pairing state unstable. In these cases, the weight of the LDOS is transferred to the region at $\omega \simeq \pm U/2$, corresponding to the characteristic energy of charge excitations at the QD (data not shown), which indicates that the QD is in the local-moment (free-spin) regime. Therefore, we conclude that as bias voltage increases, the LDOS first exhibits a crossover from the Kondo-dominant regime to the superconducting-dominant regime, and then to the local-moment regime. 
%%%%%%%%%%%%%%%%%%%%%%
%%%%%%%%%%%%%%%%%%%%%%

The above discussions on the LDOS enable us to clarify the origin of the two peaks in the differential conductance in the large-$U$ regime in Figs. \ref{fig:Udep}(a) and \ref{fig:Gsdep}(a). We first recall that the zero-bias conductance is substantially suppressed by $U$, as has been shown in Fig. \ref{fig:Udep}(a). However, for a small bias voltage, the effective cotunneling process is enhanced owing to the nonequilibrium Andreev tunneling via the Kondo resonance, which gives rise to the anomalous increase in the differential conductance at a small but finite bias voltage. Consequently, we can say that the anomalous increase found in the conductance near zero bias is a unique nonequilibrium phenomenon induced by the interplay of the Kondo effect and superconducting proximity effect. Here, we determine why the anomalous peak is located at $eV/\tilde{\Gamma}_{N} \simeq 1$. For $U=0$ in Fig.~\ref{fig:Udep}(a), the zero-bias conductance has a maximum value for $\tilde{\Gamma}_{S}/\tilde{\Gamma}_{N}=1$. Namely, the maximum transport probability is realized when the width and position of the Andreev resonance have the same value. This specific correspondence holds for the zero-bias conductance even in the interacting case because the low-energy physics of the N-QD-S system can be described by the local Fermi liquid theory.~\cite{Cuevas, Tanaka} For instance, as $\tilde{\Gamma}_{S}$ becomes smaller than $\tilde{\Gamma}_{N}$ with increasing $U$, the system loses the above specific balance, resulting in the suppression of the zero-bias conductance. Even under such conditions as $\tilde{\Gamma}_{S}/\tilde{\Gamma}_{N} \ll 1$, however, the peak position of the Andreev resonance is shifted and determined by the bias voltage $V$ instead of by $\tilde{\Gamma}_{S} $, as shown in Fig. \ref{fig:dos}. Accordingly, we can state the following by generalizing the above correspondence to the nonlinear case: In nonequilibrium cases with $eV/\tilde{\Gamma}_{N} \simeq 1$, where the position of the Andreev resonance is the same as its width, the transmission probability is enhanced, which leads to the anomalous peak of the differential conductance. In contrast, at $eV/\Gamma_{N}\simeq 0.6$, the superconducting pairing state is dominant and thus the Andreev reflection is enhanced, leading to a broad peak in the bias voltage dependence of the conductance. Namely, the two-peak structure found for the conductance in Fig.~\ref{fig:Udep}(a) clearly characterizes a crossover from the Kondo-dominant regime to the superconducting-dominant regime, and then to the local-moment regime. Even in the asymmetric-coupling cases shown in Fig. \ref{fig:Gsdep}(a), we can see the increase in conductance at a small finite bias voltage, the origin of which is the same as that in the symmetric case. On the other hand, the broad peak at higher voltages depends on the asymmetry of the couplings. For $\Gamma_{S}/\Gamma_{N}=0.5$ (=2.0), the Kondo correlation is stronger (weaker) than that in the symmetric coupling case, so that a larger (smaller) voltage is necessary to make the superconducting correlations dominant, which shifts the peak position to a higher (lower) voltage side in Fig.~\ref{fig:Gsdep}(a). The above arguments clearly explain the origin of our new findings on conductance in this paper.

%%%%%%%%%%%%%%%%%%%%%%%%%%%%%%%%%%
\begin{figure}
\centering
\includegraphics[width=7cm,clip]{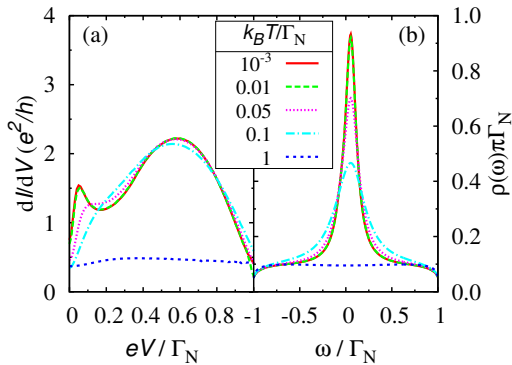}% Here is how to import EPS art
\caption{\label{fig:Tdep} (Color online) (a) Differential conductance and (b) LDOS as a function of bias voltage $V$ for several values of temperature $T$: $U/\Gamma_{N}=20$, $\epsilon_{d}/U=-0.5$, $\Gamma_{S}/\Gamma_{N}=1$, and $\Delta/\Gamma_{N}=1$.}
\end{figure}\
%%%%%%%%%%%%%%%%%%%%%%%%%%%%%%%%%%
Finally, we show the temperature dependences of the differential conductance and LDOS. Figure \ref{fig:Tdep}(a) shows differential conductance as a function of bias voltage for several temperatures $T$. It is seen that with increasing $T$, the sharp peak near $eV/\Gamma_{N}=0.05$ begins to collapse at $k_{B}T/\Gamma_{N} \simeq 0.05$, and completely disappears at $k_{B}T/\Gamma_{N}=1$. This is because the Andreev-normal cotunneling process due to the Kondo effect is suppressed at high temperatures. In order to confirm it, we also calculate the temperature dependence of the LDOS for $eV/\Gamma_{N}=0.05$, as shown in Fig. \ref{fig:Tdep}(b). We see that the Kondo resonance begins to collapse at $k_{B}T/\Gamma_{N} \simeq 0.05$, and its height for $k_{B}T/\Gamma_{N}=0.1$ becomes smaller than half of that in the low-temperature limit. Note that at this characteristic temperature $k_{B}T/\Gamma_{N}=0.05$, the conductance peak at $eV/\Gamma_{N} \simeq 0.05$ is also suppressed, ensuring that the anomalous peak in the conductance is indeed caused by the Kondo effect. These characteristic $eV$ and $k_{B}T$ correspond approximately to $\tilde{\Gamma}_{N}$. In contrast, the broad hump at $eV/\Gamma_{N} \simeq 0.6$ is less sensitive to the temperature because it is caused by the proximity effect but not by the Kondo effect. Further increase in temperature finally suppresses both the Kondo effect and the proximity effect, so that there is no resonance in the LDOS for both $eV/\Gamma_{N}=0.05$ and $0.6$ at a high temperature of $k_{B}T/\Gamma_{N}=1$. This naturally explains the decrease in the differential conductance in the entire $V$ range at higher temperatures shown in Fig. \ref{fig:Tdep}(a).

In summary, we have investigated the nonequilibrium transport in a QD system coupled to normal and superconducting leads where the Andreev reflection plays an important role. Using the modified second-order perturbation theory, we have elucidated that the differential conductance develops two characteristic peaks in its bias voltage dependence as the Coulomb interaction is enhanced. In particular, the peak near the zero-bias voltage characterizes a unique phenomenon caused by a nonequilibrium Andreev tunneling via the Kondo resonance.  It has indeed been confirmed that the sharp peak disappears with increasing temperature, reflecting the suppression of the Kondo resonance. The two-peak structure found for the voltage dependence of the conductance characterizes a crossover from the Kondo-dominant regime to the superconducting-dominant regime, and then to the local-moment regime. In real experiments, different patterns in the temperature dependence of the peaks may be observed, which can give strong evidence of the interplay between the Andreev reflection and the Kondo effect at a finite bias. We think that the transport experiments proposed here will provide an important test bed for a deeper understanding of the nonequilibrium Kondo/Andreev physics in QD systems.

%%%%%%%%%%%%%%%%%%%%%%%%%%%
\section*{Acknowledgments}
%%%%%%%%%%%%%%%%%%%%%%%%%%%
We would like to thank A.~Oguri, J.~Bauer, and R.~S.~Deacon for valuable discussions. This work was partly supported by a Grant-in-Aid from MEXT Japan (Nos. 20102008 and 21540359). Y. ~Yamada was supported by the JSPS Research Fellowships for Young Scientists, and Y.~Tanaka was supported by the Special Postdoctoral Researchers Program of RIKEN.

\end{document}